\documentclass[twoside,slac_one]{revtex4}
\usepackage{graphicx}
\usepackage{fancyhdr}
\usepackage{amsmath} 
\usepackage{bm}
\usepackage{amsxtra}
\usepackage{amssymb}
\usepackage{amsthm}
\usepackage{latexsym}
\usepackage{lscape}

\begin{document}

\title{ATLAS Experiment Pixel Detector Upgrades}

%

\author{M. Garcia-Sciveres for the ATLAS Collaboration}
\affiliation{Lawrence Berkeley National Lab, Berkeley, CA 94611}

\begin{abstract}
The ATLAS experiment pixel detector upgrade plans are focused
on development of new technology, including the FE-I4 
readout integrated circuit. The first upgrade project to 
make use of this new technology is an ``Insertable B-Layer'' (IBL) 
pixel detector to be installed on a replacement beam pipe in 2013. 
The IBL will fit inside and not alter the existing ATLAS pixel detector. 
However, the possibility is being studied to replace the pixel whole detector in 2018 with a lower mass, 
higher performance instrument based on the FE-I4 chip and new mechanical structures. 
\end{abstract}

\maketitle

\thispagestyle{fancy}


\section{Introduction}
The hybrid pixel detectors in operation at the LHC today \cite{atlaspixel,cmspixel,alicepixel} and installed around 2007 represent a great achievement in particle physics instrumentation, but have limitations in rate capability, radiation hardness, power and mass, imposed by the technology that was available at the time of their design. Work within the ATLAS collaboration has since focused on developing new technology rather than on a specific detector design. 
The developments discussed here are a new readout integrated circuit called 
FE-I4, new sensor developments, mechanical structures based on a new type of high thermal conductivity carbon foam, and electro-mechanical integrated structures. The technology developments follow a road-map inspired by the requirements of a future high luminosity LHC upgrade, but has the generic goals of reducing mass and cost while increasing rate capacity and radiation tolerance. As individual technologies mature, they can be deploy as soon as the need or opportunity arises. The ``Insertable B-Layer'' (IBL) upgrade~\cite{IBL} will deploy in 2013 the FE-I4 integrated circuit, high thermal conductivity carbon foam, "slim edge" planar sensors, and 3D sensors. Other near term upgrades are being considered that would also deploy the FE-I4 integrated circuit along with diamond sensors (a beam luminosity monitor) and 3D sensors (a far forward pixel detector). Together with further new technologies (on-stave power conversion, high speed data transmission, etc.), all the technologies described in this paper would be available for deployment on a 2018 time scale and this possibility is under consideration. 

\section{FE-I4 Integrated Circuit}

\begin{figure}[htb]
\includegraphics[width=80mm]{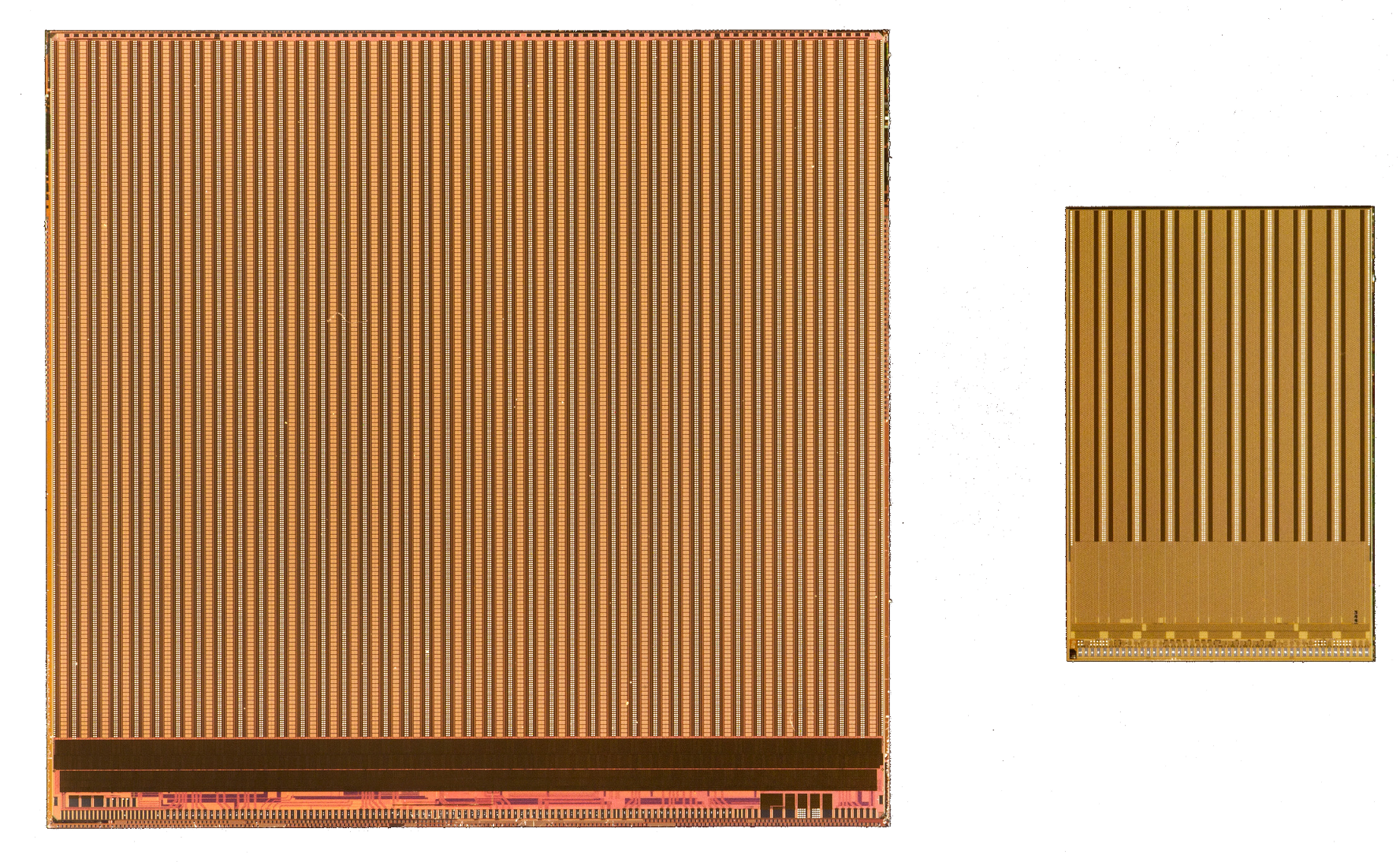}
\centering
\caption{\label{fig:chips} Photograph of the FE-I4 readout chip (left) and FE-I3 (right). The width and height as shown 
are 20.3\,mm by 19.2\,mm for FE-I4 and 
7.4\,mm by 11.0\,mm for FE-I3}
\end{figure}

The FE-I4 pixel readout integrated circuit has been developed with the generic goals of increasing rate capability and radiation tolerance and reducing system cost~\cite{FEI4}. Cost reduction is mainly achieved by making the integrated circuit area as large as possible, since flip chip bump bonding cost scales approximately with the number of chips processed rather than with area processed, and until now bump bonding has been the main cost driver. FE-I4 is therefore the largest pixel readout IC produced to date: Fig~\ref{fig:chips} shows a photograph of FE-I4 next to the present ATLAS pixel readout chip called FE-I3. Bump bonding cost reduction is not critical for the IBL project because the total active area is small (0.15\,m$^2$). However, the large format of FE-I4 is critical to fitting the tight radial envelope constraints of IBL. Additionally, only 10\% of the FE-I4 area is devoted to periphery (compared to 30\% for FE-I3) and this is again critical for the IBL geometry. 

The small FE-I4 periphery is a secondary benefit of a new readout 
architecture~\cite{architecture}. The reduced periphery is achieved because all hit storage is within the pixel matrix itself. 
The primary purpose of this new architecture is to achieve much higher rate capability than previous chips. 
This is critical for IBL (due to higher hit rate at smaller radius). 
The new readout stores hit charge and timing locally within a 4-pixel region until selected by an external trigger. This was possible by taking advantage of the high digital circuit density in the 130\,nm feature size CMOS process used.
Increased radiation tolerance is also achieved by the 130\,nm process. 
Each FE-I4 chip has 26,\,880 pixels 50\,$\mu$m wide by 250\,$\mu$m long.

\section{Sensors}

Pixel sensor development has been ongoing in several large collaborations~\cite{planar,diamond,3d}. Within ATLAS the IBL project provided a significant boost to pixel sensor prototyping and characterization. 

A planar n-in-n IBL sensor design is based on the sensors used in the present
ATLAS pixel detector~\cite{planar1}, adapted
to the IBL needs by reducing the pixel size to match FE-I4 granularity and the 
sensor wafer thickness to 200\,$\mu$m. A new guard ring design maximizes
the possible operation voltage and minimizes the inactive edge. 
The so-called``slim-edge'' design, uses 13 guard rings on the bias side of the sensor to manage the potential
drop between bias contact and physical sensor edge. The distance from the last guard to the dicing
edge is 100\,$\mu$m. The edge pixels on the electrode side of the sensor 
overlap the guard ring area to collect charge from as close as possible
to the physical edge. Further details can be found in~\cite{planar2}. 

A ``double sided'' 3D sensor IBL design has the measuring and biasing electrodes etched into the p-type silicon bulk from opposite sides, 
penetrating most of the 
thickness, but not all the way through. This makes processing and bias 
connection simpler. The sensor bulk is laterally depleted between
n$^+$-type measuring electrodes and p$^+$-type biasing electrodes. 3D sensors allow for high field strength and good charge collection
after heavy irradiation with relatively low bias voltages. The 
IBL 3D design uses two n$^+$ electrodes per pixel,
surrounded by six p$^+$ biasing electrodes, to provide good charge collection 
without excessive capacitance~\cite{3d1}. The sensors are 230\,$\mu$m thick  and have an inactive guard ring area 225\,$\mu$m wide. Two guard ring
variants have been developed using combinations of n$^+$ columns and p$^+$ columns
forming fences around the active area.

\section{Mechanics and Modular Structures}

\begin{figure}[htb]
\includegraphics[width=80mm]{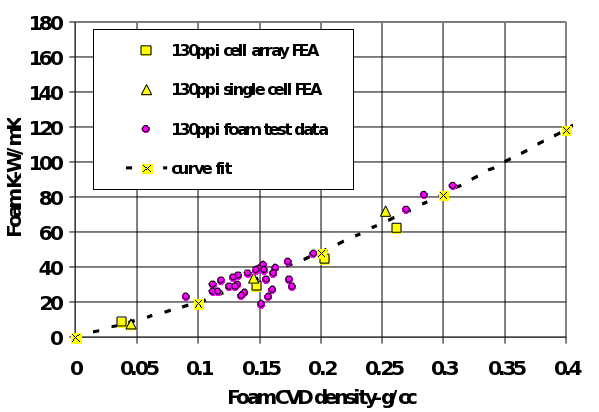}
\centering
\caption{\label{fig:foamk} Thermal conductivity vs. density for new carbon foam. Finite element calculates (squares and triangles) are compared with sample measurements (circles).}
\end{figure}

A new type of all-carbon thermally conductive foam has been developed for 
silicon pixel and strip mechanical supports~\cite{SBIR}. The foam is readily machinable and has low radiation length, yet thermal conductivity two orders of 
magnitude higher than conventional foams of the same density. Fig.~\ref{fig:foamk} shows finite element 
analysis predicted foam conductivities compared to sample measurements. 
The foam is produced using a reticulated vitreous carbon (RVC) foam precursor
and increasing the thermal conductivity through combination of chemical vapor
deposition and heat treatment. A standard fabrication has
been developed producing blocks 2.5\,cm thick and $30 \times 30$\,cm$^2$. 

A common basic concept uses a small coolant tube glued to foam, 
coupled with high stiffness carbon facings to form rigid, lightweight structures supporting modules and their electrical connections. Due to the high foam 
conductivity, such structures efficiently couple heat for small tubes typical 
of CO$_2$ evaporative cooling. This is the ATLAS baseline for all future pixel
support structures (including IBL) and also a candidate for silicon strip supports. The reliability of such foam-based structures has been validated by
extensive thermal cycling and repeated shock, before and after irradiation up to
1\,Grad. 

\begin{figure}[htb]
\includegraphics[width=135mm]{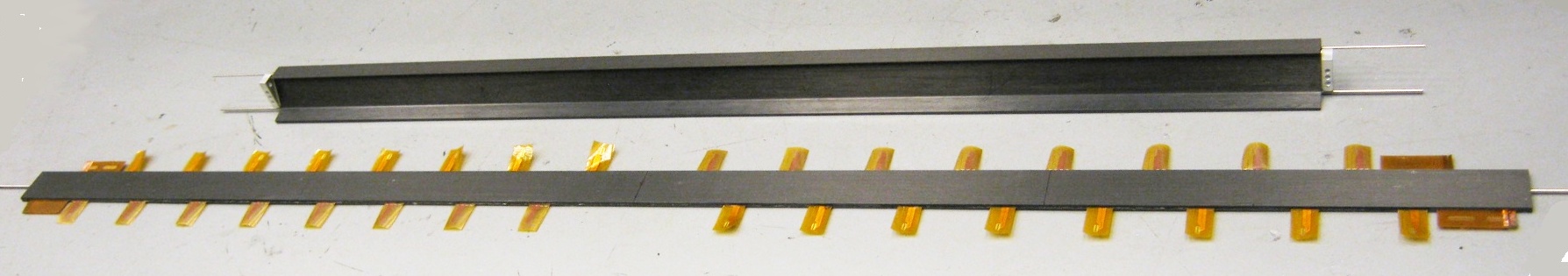}
\centering
\caption{\label{fig:staves} Photograph of a 1.4\,m long box beam 
stave prototype (bottom) 
and a 1\,m long I-beam stave prototype (top).}
\end{figure}

\begin{figure}[htb]
\includegraphics[width=80mm]{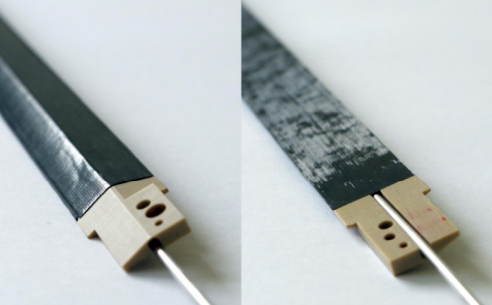}
\centering
\caption{\label{fig:iblstave} Photographs of an end of an IBL stave. The width is 2\,cm.}
\end{figure}

Three types of pixel structures have been prototyped, all based on evaporative 
CO$_2$ cooling in Ti tubes coupled to the surfaces with thermally conductive 
carbon foam. These are the IBL stave, box beam stave, and ``I-beam'' coupled layer structures. 
Box beams have been made up to 1.4\,m long by 4\,cm wide, including embedded
electrical cables. Also demonstrator bent 
sections have been made, which can be used to increase pseudorapidity coverage of barrel sections improving the transition to forward coverage. I-beam coupled
layers have been made up to 1\,m long. 
By radially coupling two measurements layers, a very stiff structure can be made, thereby saving mass in support frame structures- even long I-beams can be end-supported. Fig.~\ref{fig:staves} shows a 1.4\,m long
box beam prototype with embedded cables next to a 1\,m long I-beam prototype
Fig.~\ref{fig:iblstave} shows detail views of one end of a 2\,cm wide IBL stave.

Stave cable designs have been produced compatible with serial powering of all modules on a stave, using aluminum conductors. The FE-I4 chip contains so called
``Shunt-LDO'' voltage regulators~\cite{shuldo} designed to support serial power distribution. In IBL these regulators are used to permit safe operation with point-to-point power over resistive cables (the voltage drop in the cables is comparable to the load). 

\section{IBL and possible future detectors}

\begin{figure}[htb]
\includegraphics[width=80mm]{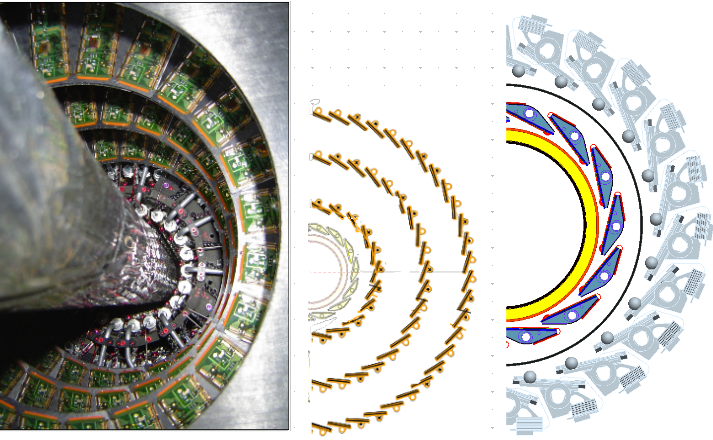}
\centering
\caption{\label{Fig:IBLphotos} Left: photograph of the present pixel detector as it is mounted around the beam pipe. Three disks are in the foreground, with the end of the barrel leaving a roughly 1\,cm gap to the beam pipe in the background. Center: section view of the three barrel layers on approximately the same scale as the photograph, together with the IBL and new beam pipe section inside. Right: detail showing only the inner layer of the present detector (gray), and IBL and new beam pipe inside. The black line separating the IBL from the present inner layer is a carbon fiber tube that will support and guide insertion of the IBL.}
\end{figure}

The Insertable B-Layer (IBL) is the first planned upgrade to the ATLAS inner tracker. It consists of new barrel layer of pixels to be mounted inside the existing detector, without altering it. The present ATLAS pixel inner layer has approximately a 1\,cm gap to the outer envelope of the beam pipe, and additional room will be created by replacing the existing beam pipe with a new one of reduced diameter. In the context of new technology development, it is interesting to recall that the original ATLAS pixel technical design report~\cite{pixelTDR} called for a smaller radius, smaller pixel size inner layer than was eventually implemented. The reason for the change was lack of sufficiently advanced technology- a smaller pixel size chip could not be designed in the CMOS process used at the time, and the radiation and rate tolerance requirements could not be met at the desired small radius. In this sense, the new technology that is being deployed in IBL can finally achieve the original ATLAS design goals. Fig.~\ref{Fig:IBLphotos} shows a photograph of the present detector exactly as it is mounted around the beam pipe, as well as cross sections view of barrel layers and the IBL. 

The IBL is approximately 64\,cm long at a mean radius of 3.2\,cm. Thanks to the small radius, a single barrel covers the full pseudorapidity range of the ATLAS tracking system. It is made up of 14 staves with 32 FE-I4 chips each. The IBL mass target in the active region is 1.5\% of a radiation length at normal incidence. The additional high precision measurement, at smaller radius and with low mass translates into a significant improvement in the performance of he ATLAS detector: the purity of b-quark jets identified by the presence of a displaced vertex can be increased by nearly a factor 
of two~\cite{IBL}. The full detector is expected to degrade due to radiation damage, random failures, and inefficiency at high interaction rate.  Even after such degradation, simulations show that the presence of the radiation hard, high rate capable IBL translates into better performance than is achieved today, before degradation, but without IBL.   

While IBL is a near term upgrade already in progress, eventually the full ATLAS tracker must be replaced to cope with a machine luminosity upgrade, and the technology development aims at this goal. However, if technology matures on a faster timescale than strictly needed for this ultimate replacement, it may be advantageous to advance the upgrade of certain elements. This early availability of new technology is precisely what has made the IBL upgrade possible on the 2013 time scale. ATLAS is now studying the potential advantages of upgrading the full pixel detector (the part outside IBL) before the LHC high luminosity upgrade. The new detector would have lower mass, higher granularity, bigger lever arm for track finding, and would already meet the rate and radiation requirements of the high luminosity LHC. The merits of such an upgrade will be assessed on two fronts: performance studies using Monte Carlo simulation to evaluate possible gains in physics reach, and projections of expected degradation of the existing detector, including random failures, radiation damage, and efficiency loss at high rate. 

\begin{figure}[htb]
\includegraphics[width=135mm]{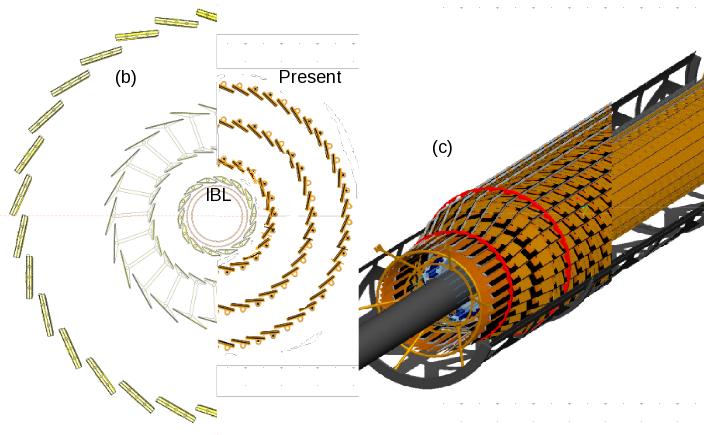}
\centering
\caption{\label{fig:newpix} Comparison of barrel cross sections for present detector with IBL and (a) potential replacement on 2018 timescale. (b) Model view of replacement detector having barrel cross section (a).}
\end{figure}

Fig.~\ref{fig:newpix} compares the barrel section of the present detector with IBL to a possible upgrade layout keeping the same IBL. This new pixel layout is based on mechanical structures already prototyped (box beam staves, bent sections, and I-beams) and on modules made with 4 FE-I4 chips. Such modules are presently under development. The new detector would contain approximately 250 million pixels in just over 3\,m$^2$ of silicon, compared to 80 million pixels in under 2\,m$^2$ for the present detector (the pixels are smaller, but also the outer layer radius is larger).

\section{Conclusion}

Development of new technologies has been made a priority of ATLAS pixel R\&D, with the 
result that several advances are ready today and will be deployed in the IBL upgrade (2013). These include the new FE-I4 readout chip, new sensors, and low mass foam. Further items are being developed and will be ready for 2018 deployment if needed. These include low mass I-beam and bent box beam stave structures with integrated services, lower cost per unit area (large format of FE-I4 chip, sensors on 6 inch wafers, faster bump bonding), low mass cables, and on-stave power conversion. Other items that will be ready for 2018 deployment, but were not covered include low mass multi GHz signal cables, high speed data transmission chips, and data acquisition. ATLAS is studying the benefits of a potential 2018 pixel upgrade taking advantage of the above new technologies. Still further advances are in early stages of development and will likely not mature for deployment before 2020. These include a further chip generation beyond FE-I4 (higher rate and radiation) even more radiation hard sensors, and very high speed communication.

\bigskip 

\end{document}